\documentclass[prl,a4paper,twocolumn,showpacs,floatfix,superscriptaddress,amsmath,amsfonts,amssymb,preprintnumbers,citeautoscript]{revtex4}
\pdfoutput=1
\usepackage[T1]{fontenc}\usepackage[latin1]{inputenc}
\usepackage{dcolumn,graphicx,color,booktabs,afterpage}
\usepackage[charter,greekuppercase=italicized]{mathdesign}
\usepackage{comment}
\renewcommand{\figurename}{Fig.}
\renewcommand{\tablename}{Table}
\makeatletter\renewcommand{\fnum@figure}[1]{\figurename~\thefigure~(color online).}\makeatother
\makeatletter\renewcommand{\fnum@table}[1]{\tablename~\thetable.}\makeatother
\newcount\hh \newcount\mm
\hh=\time \divide\hh by 60
\mm=\hh \multiply\mm by 60 \mm=-\mm
\advance\mm by \time
\def\now{\number\hh:\ifnum\mm<10{}0\fi\number\mm}

\usepackage[colorlinks,plainpages=false,linkcolor=blue,urlcolor=blue,citecolor=blue,pdfpagemode=UseNone,pdfstartview=FitBH]{hyperref}

\newcommand{\onehalf}{\frac{\mbox{\scriptsize1}}{\protect\raisebox{0pt}{\scriptsize2}}}

\newcommand{\quarter}{\frac{\mbox{\scriptsize1}}{\protect\raisebox{0pt}{\scriptsize4}}}
\newcommand{\vect}[1]{\textbf{#1}}

\newcommand{\ceb}{CeB$_6$}
\newcommand{\celab}{Ce$_{1-x}$La$_x$B$_6$}
\newcommand{\cci}{CeCoIn$_5$}
\newcommand{\yrs}{YbRh$_2$Si$_2$}
\newcommand{\yrsg}{YbRh$_2$(Si$_{1-x}$Ge$_{x}$)$_{2}$}
\newcommand{\cps}{Ce$_3$Pd$_{20}$Si$_6$}
\newcommand{\celarusi}{Ce$_{1-x}$La$_x$Ru$_2$Si$_2$}

\begin{document}

\makeatletter\renewcommand{\ps@plain}{%
\def\@evenhead{\hfill\itshape\rightmark}%
\def\@oddhead{\itshape\leftmark\hfill}%
\renewcommand{\@evenfoot}{\hfill\small{--~\thepage~--}\hfill}%
\renewcommand{\@oddfoot}{\hfill\small{--~\thepage~--}\hfill}%
}\makeatother\pagestyle{plain}


\title{Magnetic-field and doping dependence of low-energy spin fluctuations\\ in the antiferroquadrupolar compound \celab}

\author{G.~Friemel}
\affiliation{Max-Planck-Institut für Festkörperforschung, Heisenbergstraße 1, 70569 Stuttgart, Germany}

\author{H.~Jang}
\affiliation{Max-Planck-Institut für Festkörperforschung, Heisenbergstraße 1, 70569 Stuttgart, Germany}
\affiliation{Stanford Synchrotron Radiation Lightsource, SLAC National Accelerator Laboratory, Menlo Park, California 94025, USA}

\author{A.~Schneidewind}
\affiliation{J\"ulich Center for Neutron Science (JCNS), Forschungszentrum J\"ulich GmbH, Outstation at Heinz Maier\,--\,Leibnitz Zentrum (MLZ), D-85747 Garching, Germany}

\author{A.~Ivanov}
\affiliation{Institut Laue-Langevin, 6 rue Jules Horowitz, 38042 Grenoble Cedex 9, France}

\author{A.\,V.~Dukhnenko}
\affiliation{I.\,M. Frantsevich Institute for Problems of Materials Science of NAS, 3 Krzhyzhanovsky str., Kiev 03680, Ukraine}

\author{N.\,Y.~Shitsevalova}
\affiliation{I.\,M. Frantsevich Institute for Problems of Materials Science of NAS, 3 Krzhyzhanovsky str., Kiev 03680, Ukraine}

\author{V.~B.~Filipov}
\affiliation{I.\,M. Frantsevich Institute for Problems of Materials Science of NAS, 3 Krzhyzhanovsky str., Kiev 03680, Ukraine}

\author{B.~Keimer}
\affiliation{Max-Planck-Institut für Festkörperforschung, Heisenbergstraße 1, 70569 Stuttgart, Germany}

\author{D.~S.~Inosov}
\affiliation{Institut für Festkörperphysik, TU Dresden, D-01069 Dresden, Germany}

\begin{abstract}

\noindent \ceb\ is a model compound exhibiting antiferroquadrupolar (AFQ) order, its magnetic properties being typically interpreted within localized models. More recently, the observation of strong and sharp magnetic exciton modes forming in its antiferromagnetic (AFM) state at both ferromagnetic and AFQ wave vectors suggested a significant contribution of itinerant electrons to the spin dynamics. Here we investigate the evolution of the AFQ excitation upon the application of an external magnetic field and the substitution of Ce with non-magnetic La, both parameters known to suppress the AFM phase. We find that the exciton energy decreases proportionally to $T_\text{N}$ upon doping. In field, its intensity is suppressed, while its energy remains constant. Its disappearance above the critical field of the AFM phase is preceded by the formation of two modes, whose energies grow linearly with magnetic field upon entering the AFQ phase. These findings suggest a crossover from itinerant to localized spin dynamics between the two phases, the coupling to heavy-fermion quasiparticles being crucial for a comprehensive description of the magnon spectrum.
\end{abstract}

\pacs{75.30.Mb, 75.30.Ds, 78.70.Nx\vspace{-5pt}}

\maketitle\enlargethispage{3pt}

A current focus of research in heavy fermion (HF) compounds is the study of quantum critical points (QCP) --- phase transitions achieved at zero temperature by tuning an external parameter such as magnetic field, doping, or pressure. One possible signature of a QCP is the change of the quasiparticle character from localized to itinerant, when the transition is connected with a breakdown of the Kondo effect and the removal of $f\!$-electrons from the Fermi surface (FS). Such an effect was observed, for example, by transport measurements in the prototypical QCP system \yrsg\ at the critical field of the low-temperature antiferromagnetic (AFM) phase \cite{CustersGegenwart03, Paschen2004}. Recently, the list of QCP materials was extended with the cubic Kondo lattice compound \cps\ \cite{CustersLorenzer12, PortnichenkoCameron15, YamaokaSchwier15}, whose magnetic phase diagram comprises an antiferroquadrupolar (AFQ) phase below $T_\text{Q}=0.5\,\mathrm{K}$ and an AFM phase at even lower temperatures. For the latter phase, a field-induced QCP was observed at the critical field $B^{*}=0.9\,\mathrm{T}$ and the concomitant FS reconstruction was related to the destruction of the Kondo effect \cite{CustersLorenzer12}.

\begin{figure}[b!]\vspace{-7pt}
\includegraphics[width=0.9\columnwidth]{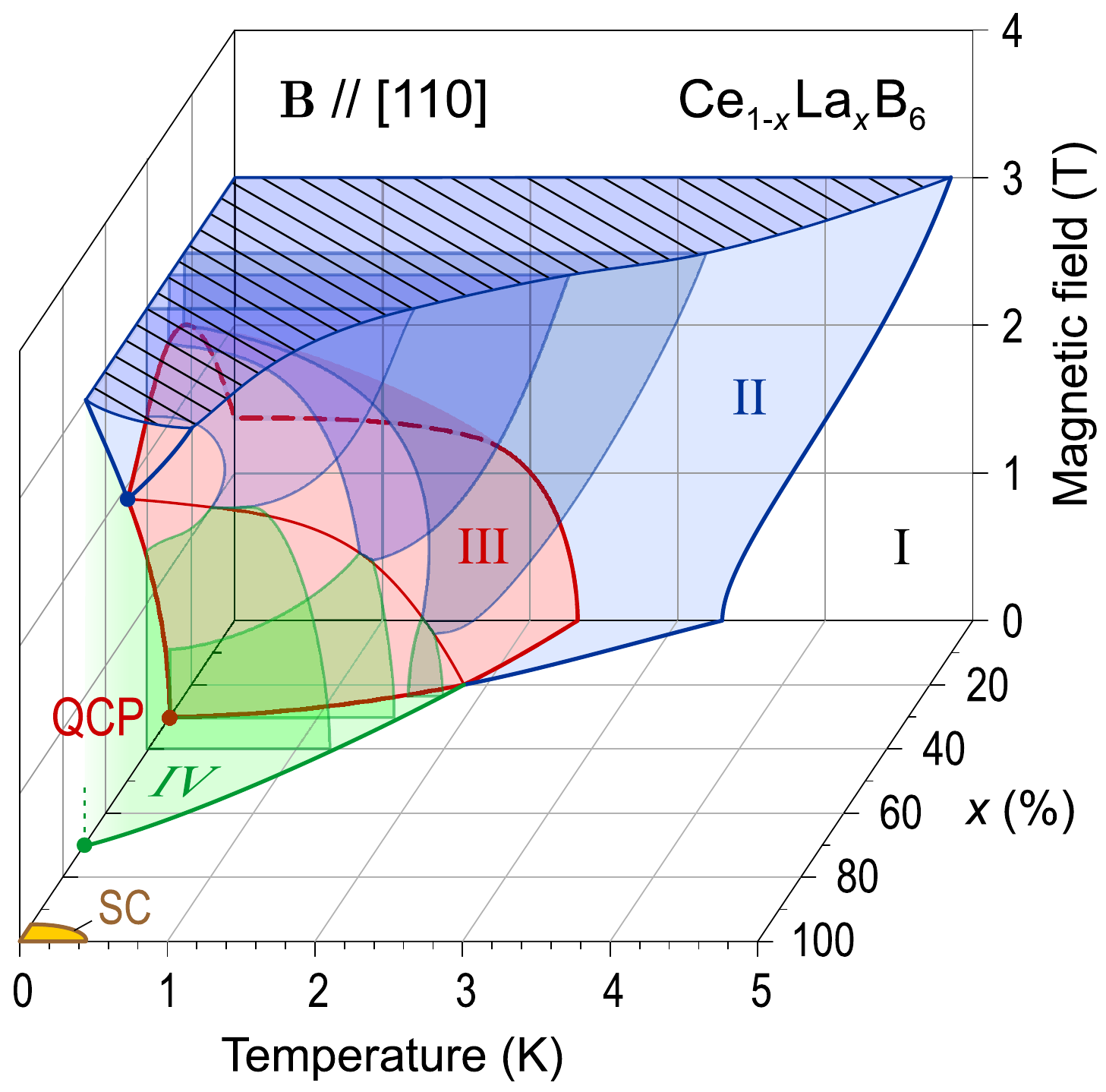}
\caption{The phase diagram of Ce$_{1-x}$La$_x$B$_6$ vs. temperature, La doping level, and magnetic field applied along the [110] crystallographic direction, reconstructed from the data in Refs.~\citenum{HiroiKobayashi97, TayamaSakakibara97, HiroiKobayashi98, KobayashiSera00}. The following phases are marked by color in the online version of the article: paramagnetic (phase I, clear), AFQ (phase II, blue), AFM (phases III/III$^\prime$, red), antiferrooctupolar (phase IV, green). The small superconducting dome of LaB$_6$ is schematically shown at the bottom.}
\label{fig1}\vspace{-3pt}
\end{figure}

\begin{figure*}[t]\vspace{-1.2em}
\includegraphics[width=\textwidth]{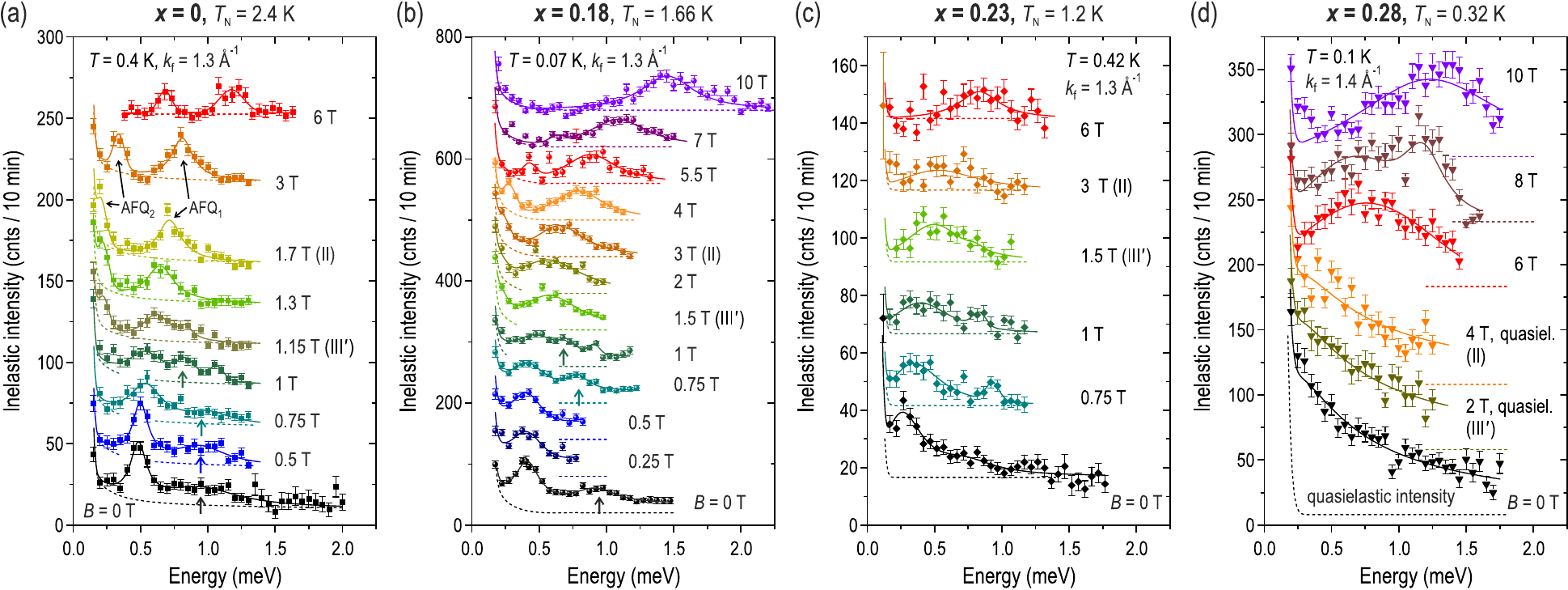}
\caption{INS spectra measured at the AFQ wave vector, $R(\frac{1}{2}\,\frac{1}{2}\,\frac{1}{2})$, in different magnetic fields $\vect{B}\parallel (1\overline{1}0)$ for (a) $x=0$, (b) $x=0.18$, (c) $x=0.23$, (d) $x=0.28$. The spectra are shifted vertically for clarity, the dashed lines indicating the background baseline for each spectrum. Solid lines represent fits described in the text.}
\label{fig2}\vspace{-0pt}
\end{figure*}

\ceb\ was one of the first known AFQ compounds, where the multipolar order was observed both indirectly as an anomaly in specific heat at $T_\text{Q}=3.2\,\mathrm{K}$ \cite{FujitaSuzuki80} and directly by resonant x-ray diffraction \cite{CeB6resonantXray} or neutron diffraction \cite{EffantinRossat-Mignod85, FriemelLi12} as a weak magnetic Bragg peak centered at the $\vect{Q}_\text{AFQ}=R(\onehalf\onehalf\onehalf)$ propagation vector. The magnetic phase diagram of \ceb\ \cite{EffantinRossat-Mignod85} is similar to that of \cps, yet with larger temperature and magnetic-field scales. Correspondingly, it features an AFM phase (phase III) below $T_\text{N}=2.4\,\mathrm{K}$, which exhibits a complex double-$\vect{q}$ structure \cite{BurletRossatMignod82+ZaharkoFischer03} with $\vect{q}_1={\mit\Sigma}(\quarter\,\quarter\,0),\,\vect{q}_1^{\prime}=S(\quarter\,\quarter\,\onehalf)$ and $\vect{q}_2={\mit\Sigma}_2(\quarter\,\overline{\quarter}\,0), \vect{q}_2^{\prime}=S_2^{\prime}(\quarter\,\overline{\quarter}\,\onehalf)$. The AFM phase can be suppressed in a magnetic field of $B_\text{c}=1.05\,\mathrm{T}$ \cite{EffantinRossat-Mignod85}, however, in contrast to \cps, resistivity and heat capacity exhibit Fermi-liquid-like behavior down to the lowest temperatures \cite{NakamuraEndo06, Bredl87}, suggesting the absence of field-induced quantum-critical fluctuations. Instead, CeB$_6$ enters an intermediate magnetic phase (phase III$^{\prime}$) for $B_\text{c}<B<B_\text{Q}$ \cite{EffantinRossat-Mignod85,KunimoriKotani11}. For $B>B_\text{Q}=1.7\,\mathrm{T}$, the AFQ phase is established and stabilized up to very high fields, showing an increase of $T_\text{Q}$ vs. $\!B$ \cite{FujitaSuzuki80,EffantinRossat-Mignod85}. Beside magnetic field, substitution with non-magnetic lanthanum in \celab\, also leads to a suppression of the AFM phase with a critical doping level $x_\text{c}=0.3$ \cite{HiroiKobayashi97, TayamaSakakibara97}. However, the transition at zero temperature occurs into an enigmatic phase IV \cite{HiroiKobayashi98, KobayashiSera00, KobayashiYoshino03} instead of the paramagnetic phase (see Fig.~\ref{fig1}), also precluding the direct observation of quantum-critical fluctuations in transport properties \cite{NakamuraEndo06}. The $B$\,--\,$T$ phase diagram as well as the temperature and magnetic-field dependencies of the uniform and staggered magnetization could be successfully modeled by a purely localized mean-field Hamiltonian consisting of Zeeman, dipolar, quadrupolar and octupolar exchange terms \cite{SeraKobayashi99+SeraIchikawa01+ShiinaSakai98}, which suggested that \ceb\ lies far from the critical point where the Kondo effect breaks down.

However, this localized viewpoint has been challenged by recent inelastic neutron scattering (INS) experiments, demonstrating the appearance of a sharp resonant mode at $\vect{Q}_\text{AFQ}$, centered at an energy $\hbar\omega_R=0.5\,\mathrm{meV}$, in the AFM phase \cite{FriemelLi12}. It can be explained as a pole in the itinerant spin susceptibility calculated in the random-phase-approximation (RPA) for the HF ground state \cite{AkbariThalmeier12}, signifying a close relationship to the sharp resonant modes observed in the superconducting (SC) state of some other HF compounds, such as CeCoIn$_5$ \cite{StockBroholm08, EreminZwicknagl08}, CeCu$_2$Si$_2$ \cite{StockertArndt11}, or the antiferromagnetic superconductor UPd$_2$Al$_3$ \cite{SatoAso01, BlackburnHiess06, ChangEremin07}. Such sharp magnetic excitations of itinerant origin, which are usually well localized both in energy and momentum, are referred to as spin excitons to be distinguished from conventional magnons (spin waves) or crystal-field excitations in localized magnets. Since the excitonic origin of the $R$-point resonant mode in CeB$_6$ has been suggested earlier \cite{FriemelLi12, AkbariThalmeier12}, we will stick to this terminology in the following. Later it was also established that the $R$-point exciton is connected to a ferromagnetic collective mode, which is much more intense than the spin waves emerging from $\vect{q}_1$ and $\vect{q}_1^{\prime}$, putting \ceb\ close to a ferromagnetic instability \cite{JangFriemel14}. In an attempt to differentiate between itinerant and localized descriptions of the spin dynamics, here we study the evolution of the exciton mode at $R(\onehalf\onehalf\onehalf)$ upon the suppression of the AFM state by (i) dilution with non-magnetic La$^{3+}$ in \celab\, and (ii) by the application of an external magnetic field.

\begin{figure}[b]
\includegraphics[width=\columnwidth]{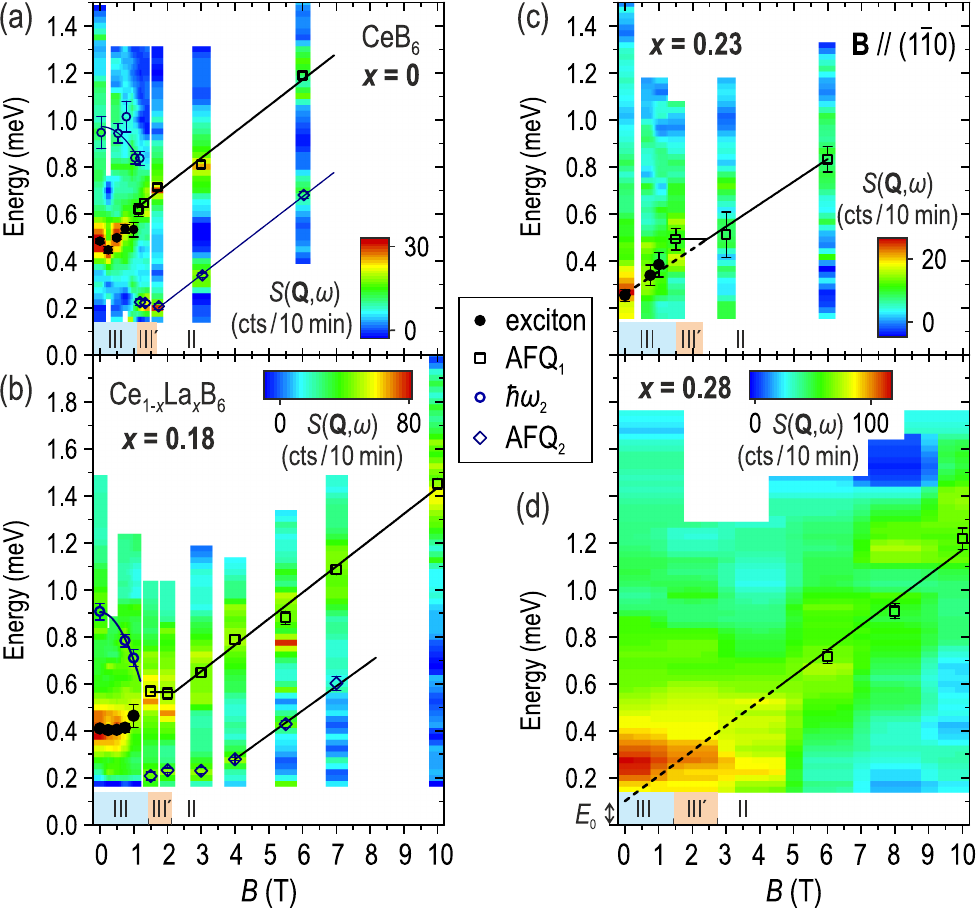}
\caption{(a)--(d) Color maps of the background-corrected intensity, $S(\mathbf{Q}_\text{AFQ},\omega)$, in the covered regions of the $\hslash\omega$-$B$ space, obtained from the data in Fig.~\ref{fig2} for (a) \ceb\ and (b)--(d) \celab\ with doping levels indicated in each panel. The intensity has been smoothed in order to decrease statistical noise and enhance readability. The symbols denote energies of the excitations derived from Lorentzian fits. The solid lines are fits described in the text.}
\label{fig3}\vspace{-3pt}
\end{figure}

\begin{figure*}[t]\vspace{-3pt}
\includegraphics[width=0.6\textwidth]{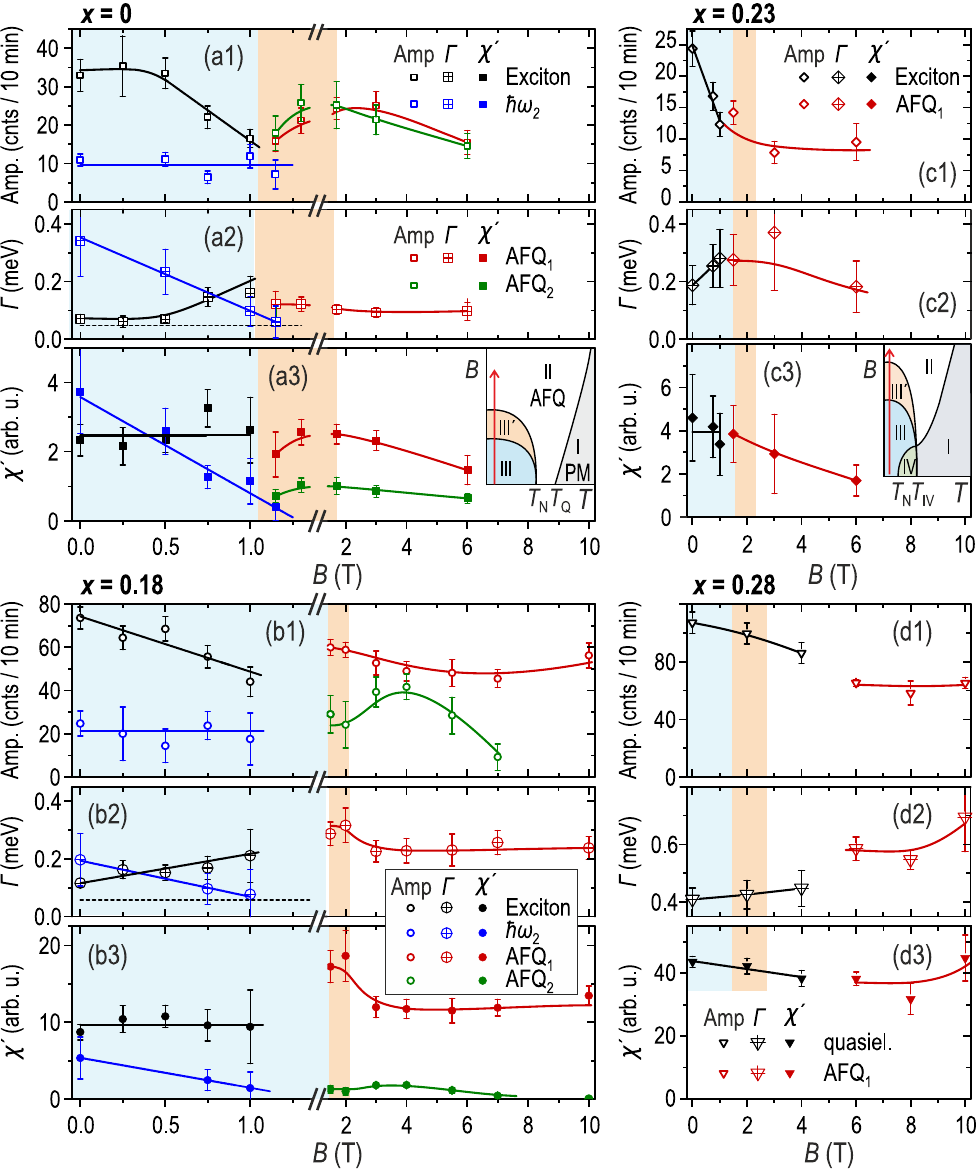}
\caption{(a)--(d) Magnetic field dependence of the amplitude [(a1)--(d1)], HWHM (${\mit \Gamma}$) [(a2)--(d2)], and the area ($\propto\chi^{\prime}$) [(a3)--(d3)] of the Lorentzian fits to the spectra shown in Fig.~\ref{fig2} for (a) \ceb, (b) $x=0.18$, (c) $x=0.23$, (d) $x=0.28$. The shaded area denotes the phases according to the phase diagram sketched in the inset to panel (a3) for $x<0.2$ and panel (c3) for $x>0.2$. The arrow denotes the position of the measurements in $B-T$ phase diagram. The low-field data ($B<6\,\mathrm{T}$) for the $x=0.28$ sample were determined from a fit to the quasielastic line shape. Solid line are guides to the eyes.\vspace{-5pt}}
\label{fig4}
\end{figure*}

We prepared rod-shaped single crystals of \celab\ ($x$~=~0, 0.18, 0.23 and 0.28), grown by floating-zone method as described elsewhere \cite{FriemelLi12}. We used 99.6\,\% isotope-enriched $^{11}$B powder as a source material to decrease neutron absorption by the $^{10}$B isotope. INS experiments were performed at the cold-neutron triple-axis instruments PANDA (MLZ, Garching, Germany) and IN14 (ILL, Grenoble, France). We fixed the final wave vector of the neutrons to $k_\text{f}=1.3$ or 1.4\,\AA$^{-1}$ for a better energy resolution and used a Be filter to suppress contamination from higher-order neutrons. The sample environment comprised a dilution or $^{3}$He insert in combination with a vertical-field magnet with the field pointing along the $[1\,\overline{1}\,0]$ direction of the crystal. The AFM transition temperatures of our samples are given in Fig.~\ref{fig2}. We determined them together with the AFQ transition temperatures, $T_\text{Q}$, and the transition fields, $B_\text{c}$ and $B_\text{Q}$, using elastic neutron scattering. The obtained values agree with the phase diagrams documented in literature \cite{FurunoSato85, HiroiKobayashi97, HiroiKobayashi98, TayamaSakakibara97, KobayashiYoshino03, KobayashiSera00, SuzukiGoto98}. Figure~\ref{fig2} shows the spectra at the $R(\onehalf\onehalf\onehalf)$ point for each sample, measured at low temperature ($T\ll T_\text{N}$) in different magnetic fields. In zero field (black curves), the $x=0$, $x=0.18$, and $x=0.23$ doped samples exhibit the exciton at $\hbar\omega_R=0.48$, $0.41$ and $0.25\,\mathrm{meV}$, respectively. In addition to the decrease in energy, the peak also broadens upon doping. Consequently, for $x=0.28$, only a quasielastic line shape,
\begin{equation}\label{Eq:QuasielasticLorentzian}
\chi^{\prime}(1-{\rm e}^{-\hslash\omega/k_{\rm B}T})^{-1}\hbar\omega{\mit \Gamma}_\text{0}/[(\hbar\omega)^2+{\mit \Gamma}_\text{0}^2],
\end{equation}
where the half-width ${\mit \Gamma}_0$ represents the relaxation rate, can be observed at low fields. Another, much broader peak $M_2$, indicated by arrows, can be seen near $\hbar\omega_2=0.94\,\mathrm{meV}$ for the $x=0$ and $x=0.18$ samples. It is worth noting that for the $x=0.28$ sample, the Bose factor leads to an asymmetric shape of the quasielastic line, which consequently exhibits a maximum at a finite energy as can be seen in the background-subtracted data [see Fig.~\ref{fig3}\,(d)]. Nevertheless, the line shape at this doping is well consistent with a quasielastic Lorentzian given by Eq.~(\ref{Eq:QuasielasticLorentzian}), which is centered at zero energy.

\begin{figure*}[t!]
\mbox{\includegraphics[width=0.7\textwidth]{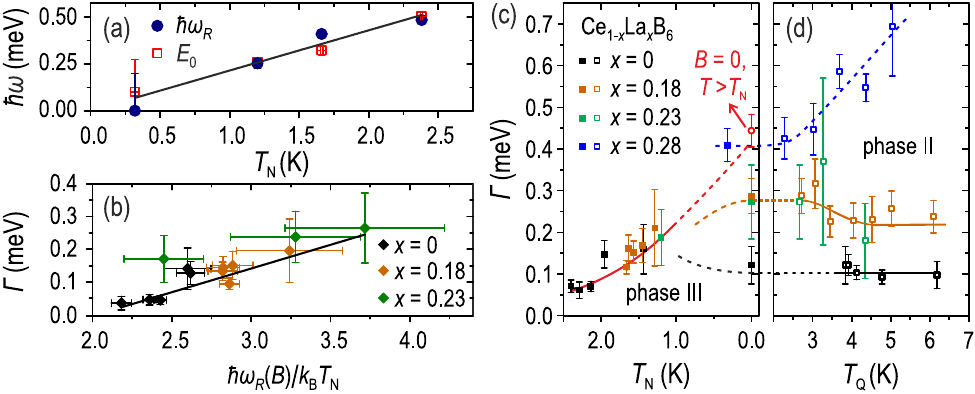}}
\caption{(a) Zero field exciton energy, $\hbar\omega_R$, and zero-field extrapolation of the AFQ$_1$ mode, $E_0$, as a function of $T_\text{N}$. (b) HWHM of the exciton, ${\mit \Gamma}$, plotted vs. $\hbar\omega_R/k_\text{B}T_\text{N}$. (c) The same vs. $\!T_\text{N}(B)$ in the AFM phase for all doping levels. Note the inverted direction of the horizontal axis. (d) ${\mit \Gamma}$ vs. $\!T_\text{Q}$ for the AFQ$_1$ mode in the AFQ phase for all doping levels. The field-dependent transition temperatures $T_\text{N}(B)$ and $T_\text{Q}(B)$ were determined from measurements of specific heat or from interpolation of the phase diagrams in literature ($x=0$, $x=0.2$, $x=0.25$) \cite{EffantinRossat-Mignod85, KobayashiSera00, KobayashiYoshino03, SuzukiNakamura05}. $T_\text{N}(B)$ for the $x=0.18$ sample was estimated from the AFM charge gap $\hbar\omega_2$. All lines are guides to the eyes.\vspace{-5pt}}
\label{fig5}
\end{figure*}

In order to analyze the complex field dependence of the spectra, we present color maps of the scattering function $S(\mathbf{Q}_\text{AFQ},\omega)$ in the measured regions of the $\hbar\omega$-$B$ parameter space  in Fig.~\ref{fig3}, where the background intensity, as given by the dashed lines for each spectrum in Fig.~\ref{fig2}, has been subtracted from the data. The field ranges of phases III, III$^{\prime}$ and II are indicated at the bottom of each panel. Taking into account the Bose factor, $\chi^{\prime\prime}(\omega)=(1-e^{-\hbar\omega/k_\text{B}T})\,S(\mathbf{Q}_\text{AFQ},\omega)$, we fitted the excitations in Fig.~\ref{fig2} to a Lorentzian line shape:
\begin{align}
\hspace{-5pt}\chi^{\prime\prime}(\omega)=\chi^{\prime}\!{\mit \Gamma}\left[\frac{1}{\hbar^2(\omega\!-\!\omega_0)^2+{\mit \Gamma}^2}-\frac{1}{\hbar^2(\omega\!+\!\omega_0)^2+{\mit \Gamma}^2}\right]
\end{align}
Here, the parameter ${\mit \Gamma}$, which describes the damping of the mode, equals the half width at half maximum (HWHM) of the peak [Fig.~\ref{fig4}\,(a2)--(d2)]. The susceptibility  $\chi^{\prime}$  is proportional to the integrated intensity (area) [Fig.~\ref{fig4}\,(a3)--(d3)]. Furthermore, the amplitude vs. $\!B$ is shown in Fig.~\ref{fig4}\,(a1)--(d1), and the mode energy $\hbar\omega_0$ vs.~$B$ is overlayed in Fig.~\ref{fig3}. In the following, we will show that the field dependence of the spin excitations can be classified according to the field regimes as outlined in the inset of Fig.~\ref{fig4}\,(a3). In the AFM phase, the exciton energy stays nearly constant vs. $\!B$, see Fig.~\ref{fig3}\,(a) and Fig.~\ref{fig3}\,(b), while its amplitude in Fig.~\ref{fig4}\,(a1) and Fig.~\ref{fig4}\,(b1) shows a strong suppression. This contrasts with the resonant mode in the SC state of \cci, whose energy splits in magnetic field with the main part of the spectral weight carried by the lower Zeeman branch \cite{StockBroholm12}. For neither of the modes do we observe any splitting in magnetic field, which agrees with the complete lifting of the degeneracy within the $\Gamma_8$ quartet ground state by the consecutive AFQ and AFM orderings. However, the energy of the high-energy mode $\hbar\omega_2$ [empty circles in Figs.~\ref{fig3}\,(a) and (b)] diminishes with field with a varying slope between the $x=0$ and $x=0.18$ compounds and a rather concave order-parameter-like field dependence. It also gets sharper, as reflected in the decreasing ${\mit \Gamma}(B)$ dependence in Figs.~\ref{fig4}\,(a2) and (b2). These facts together with the vanishing of the mode above $T_\text{N}$ let us conclude that it might correspond to the onset of the particle-hole continuum at twice the AFM charge gap. Its magnitude of $\hbar\omega_2=(0.94\pm 0.07)\,\mathrm{meV}$ in zero field for \ceb\ agrees with the $\vect{Q}$-averaged gap size of $2\Delta_\text{AFM}\approx1.2$~meV determined by point-contact spectroscopy \cite{PaulusVoss85}.

The integrated spectral weight of the exciton, corresponding to the area of the peak, remains constant with field below $T_\text{N}$ as seen in Fig.~\ref{fig4}\,(a3)--(c3) for $x=0,\,0.18$ and $x=0.23$. However, the increase of damping with field [${\mit \Gamma}$ in Fig.~\ref{fig4}\,(a2)--(c2)] reduces its amplitude. When the system enters the aforementioned phase III$^{\prime}$ above $B_\text{c}$, the amplitude starts increasing, for $x=0$ and $x=0.18$ in Fig.~\ref{fig2} (a) and (b). The peak position in energy is changing abruptly [Fig.~\ref{fig3}\,(a) and (b)] or continuously [Fig.~\ref{fig3}\,(c)]. Upon eventually entering the AFQ phase, the excitation starts shifting to higher energies, as seen in the high-field spectra for $B>2\,\mathrm{T}$ in Fig.~\ref{fig2}\,(a)--(d). Even for the $x=0.28$ sample, a rather broad mode emerges for fields $B>6\,\mathrm{T}$. This mode (we will denote it here as AFQ$_{1}$) is dominating the spectrum in the AFQ phase for all samples and has been previously observed in \ceb\ \cite{Bouvet93}. Its peak intensity [Fig.~\ref{fig4}\,(a1)--(c1)], integrated intensity [Fig.~\ref{fig4}\,(a3)--(d3)] and ${\mit \Gamma}$ [Fig.~\ref{fig4}\,(a2)--(d2)] (red points) change rather continuously when crossing the III$^{\prime}$-II phase boundary at $B_\text{Q}$ and remain nearly constant in the AFQ regime.

Moreover, upon entering phase III$^\prime$ at $B_\text{c}$, we observe the appearance of a previously unknown second mode, which can be seen for the $x=0$ and $x=0.18$ compounds at a lower energy of $\sim$\,0.2~meV in Fig.~\ref{fig2}\,(a) and (b). This excitation, denoted here as AFQ$_2$, is very sharp and evolves smoothly into the phase II [see Fig.~\ref{fig4}\,(a1) and (a3)], its energy increasing parallel to that of the AFQ$_1$ mode, as seen in Fig.~\ref{fig3}\,(a) and (b). Clarification of the nature of this new mode is left for future studies, but one can already conclude that phase III$^\prime$ and phase II are very similar in terms of spin dynamics. The linear monotonic increase of both the AFQ$_1$ and AFQ$_2$ mode energies with magnetic field in phase II (Fig.~\ref{fig3}) have a common slope $g=(0.11\pm0.004)\,\mathrm{meV/T}=(1.90\pm0.07)\mu_\text{B}$, which is doping independent. This can be qualitatively explained by a transition between two Zeeman-split energy levels, consistent with the purely localized description of the spin dynamics in a mean-field model of ordered multipoles in magnetic field \cite{ThalmeierShiina98}. For comparison, electron spin resonance (ESR) measurements, which probe the modes at the Brillouin zone center, gave a value of $g\approx1.6$ \cite{DemishevSemeno09}. The localized model would also naturally explain the increasing line width ${\mit \Gamma}$ of the AFQ$_1$ mode with La doping, shown in Fig.~\ref{fig5}\,(d), since the La-substitution alters the environment of the Ce$^{3+}$ ion, composed of six nearest neighbors.

Thus, it remains to be clarified how the exciton and the AFQ$_1$ mode are related. One possible scenario \cite{AkbariThalmeier12} describes the exciton as a collective mode below the onset of the particle-hole continuum at $2\Delta_\text{AFM}$. An alternative approach would understand the exciton as a multipolar excitation, which is overdamped by the coupling to the conduction electrons in the AFQ state $T>T_\text{N}$, but emerges as a sharp peak in the AFM state where the damping is removed by the opening of a partial charge gap \cite{FriemelLi12, JangFriemel14}. On the one hand, it would be an oversimplification to identify the exciton with the AFQ$_1$ mode, according to the second scenario, since the field dependence of the energy and the amplitude is completely different for both excitations (Figs. \ref{fig3} and \ref{fig4}). On the other hand, the zero-field extrapolation of the AFQ$_1$ mode energy $E_0$ almost coincides with the exciton energy $\hbar\omega_R$ [see Fig.~2\,(d)], both following the suppression of the magnetic energy scale, $k_\text{B}T_\text{N}$, as shown in Fig.~\ref{fig5}\,(a).

Another piece of information is given by the doping and field dependencies of the exciton line width, ${\mit \Gamma}$. Figure \ref{fig5}\,(b) shows that it increases with the ratio of the exciton energy to the AFM ordering temperature, $\hbar\omega_R/ k_\text{B}T_\text{N}$, which can be considered as a rough measure of the relative distance between the exciton and the onset of the particle-hole continuum. The points for all samples in which the exciton has been observed appear to fall on the same line, indicating that proximity to the continuum dominates the mode damping. A similar picture is given in Fig.~\ref{fig5}\,(c), where the line width is plotted directly vs. $T_\text{N}$, whose dependence on the magnetic field has been taken into account. The universality of these dependencies among all measured samples suggests that the suppression of the AFM order and the associated closing of the partial charge gap leads to a broadening of the exciton rather than the chemical disorder associated with La substitution. This ultimately leads to a quasielastic line shape in the limit of the absent phase III in zero field, reached either by temperature for $T>T_\text{N}$ (point indicated by an arrow) or by doping (for $x=0.28$), resulting in identical line widths for both cases within the measurement accuracy. In contrast, the line width of the AFQ$_1$ mode in phase II is independent of the respective AFQ energy scale, $k_{\rm B}T_\text{Q}$, as shown in panel (d). The line widths for $x=0.18$ and $x=0.23$ are comparable, which can be explained with the similar disorder effect because of chemical substitution. Were the AFQ$_1$ mode and the exciton of the same origin, we would expect a more similar response to disorder for both. Instead, the line width is smaller for the exciton and decreasing towards smaller fields, as best shown for the $x=0.18$ sample in Fig.~\ref{fig4}\,(b2). Therefore, the exciton must be derived from itinerant HF quasiparticles that are not as sensitive to the randomized local molecular field of the Ce$^{3+}$ ion as the localized AFQ$_1$ mode. The contrasting field dependencies for the energies for the exciton and the AFQ$_1$ mode in Fig.~\ref{fig3} further substantiate this conclusion.

The field-induced itinerant-localized crossover, similar to the one observed here, or metamagnetic transition points represent a topic of active research among HF compounds \cite{FlouquetAoki10}. This kind of transition was reported, for instance, in \celarusi\ ($B_\text{m}\kern-.5pt=\kern-.5pt7.7\,\mathrm{T}$ for $x=0$) \cite{AokiUji93}. The field dependence of the exciton-mode energy in Ce$_{1-x}$La$_x$B$_6$ is also analogous to that observed in UPd$_2$Al$_3$ nearly a decade ago \cite{BlackburnHiess06}. There, a sharp magnetic resonant mode associated with the superconducting state was found at a rather low energy of 0.35~meV, which could be continuously suppressed by magnetic field with only a minor softening of the peak position upon approaching the upper critical field, $H_{\rm c2}$. At higher fields, representing the normal state, a much broader inelastic peak was observed whose energy increased quasilinearly with the applied field. Similarly to our interpretation in the present work, this higher-energy inelastic feature was explained in a localized scenario as a magnetic excitation developing in a crystalline-electric-field scheme, whereas the low-energy pole in the superconducting state was later interpreted as a spin exciton within a model that took into account the dual localized-itinerant nature of the 5$f$ electrons \cite{ChangEremin07}.

In undoped or lightly doped Ce$_{1-x}$La$_x$B$_6$, the vanishing of the enigmatic mode $\hbar\omega_2$, which we suspect to be associated with the partial charge gap in the AFM phase, signals a concomitant FS reconstruction. Unlike the field-induced QCP in \yrs\ \cite{CustersGegenwart03} and \cps\ \cite{CustersLorenzer12}, no critical fluctuations are observed at either $B_\text{c}$ or $B_\text{Q}$. This can be understood considering that the itinerant magnetic moments are ferromagnetically coupled \cite{JangFriemel14}, and magnetic field stabilizes the associated spin dynamics, including the exciton. Since the AFQ order in CeB$_6$ is promoting ferromagnetism \cite{JangFriemel14,DemishevSemeno09,KobayashiSera99}, there must be a close relationship with the HF quasiparticles, which has not been taken into account in current theories \cite{SeraKobayashi99+SeraIchikawa01+ShiinaSakai98}. The role of the AFM ordering in the present interpretation is to reduce the scattering by the local 4$f\!$-spins, which enables the observation of the exciton in the first place \cite{AkbariThalmeier12}.

The substitution with La, as the second tuning parameter, suppresses both the AFM and AFQ order, reflected in a decrease of $T_\text{Q}$ and $T_\text{N}$ reaching a QCP close to $x_\text{c}=0.3$, where the AFM phase vanishes in zero field \cite{TayamaSakakibara97}. However, unlike in the conventional QCP scenario, the transition does not occur into the paramagnetic phase, but into another less studied phase IV ($T_\text{IV}=1.4\,\mathrm{K}$), as shown in the inset to Fig.~\ref{fig4}\,(c3). This phase can be induced by field starting from the doping level of $x=0.2$ \cite{KobayashiYoshino03}, yet its order parameter remains unknown, and dipolar short-range correlations coexisting with antiferrooctupolar ordering along $R(\onehalf\onehalf\onehalf)$ were proposed \cite{MannixTanaka05+TakagiwaOhishi02}. The clear transition at $T_\text{IV}$ in specific heat \cite{FurunoSato85, NakamuraGoto00} suggests that phase IV takes over the role of the low-temperature ordered phase \cite{TayamaSakakibara97}. The associated spin excitations, presented here, are quasielastic in zero field, with the AFQ$_1$ mode emerging in finite fields, where it is characterized by a significantly increased line width compared to lower doping levels [see Fig.~\ref{fig2}\,(d), Fig.~\ref{fig3}\,(d) and Fig.~\ref{fig5}\,(d)]. This could denote the onset of critical fluctuations, which arise from the suppression of the exciton energy $\hbar\omega_R$ close to zero in the $x=0.28$ sample. The same applies presumably to the ferromagnetic mode, which together with the exciton and spin-wave excitations are forming the dominant thermodynamic critical fluctuations in \ceb\ above $T_\text{N}$ \cite{JangFriemel14}. As the $E_0$ energy scale of the AFQ$_1$ mode also vanishes [Fig.~\ref{fig5}\,(a)], one can regard the AFM QCP here as coincident with the zero-field-extrapolated QCP of the AFQ phase. This QCP may also explain an enhancement of the effective mass upon approaching $x_\text{c}$ as observed in transport \cite{NakamuraEndo06}.

In conclusion, we reported the magnetic-field and doping dependence of the spin-excitation spectrum at the exciton wave vector. We demonstrated that the exciton mode of itinerant origin transforms into a localized Zeeman-type mode above the critical field $B_\text{c}$, which cannot be fully understood within the available multipolar models of the spin dynamics \cite{ThalmeierShiina98}. Contrary to the cases of \cps\ and \yrs, these fluctuations do not become critical at $B_\text{c}$, however, they are critically softened upon doping, indicating a QCP near $x_\text{c}=0.3$, which is hidden inside the enigmatic phase IV. These results outline rich prospects in the research of competing correlated ground states in the structurally simple three-dimensional system \ceb.

We thank A. Akbari, G. Jackeli, Yuan Li, and J.-M. Mignot for stimulating discussions. D.\,S.\,I. acknowledges financial support by the German Research Foundation (DFG) under grant No.~IN\,209/3-1. H.\,J. was supported by the Max Planck POSTECH Center for Complex Phase Materials with KR2011-0031558.

\end{document}